\documentclass[a4paper,12pt,fleqn]{article}
\usepackage{latexsym}

\usepackage{amssymb}
\usepackage{amsmath}
\usepackage{amsfonts}

\font\helv=cmssbx10

\def\brho{\pmb{\rho}}
\def\beeta{\pmb{\eta}}

\def\bdelta{\pmb{\delta}}
\begin{document}
\title{\bf Pseudo-unitary symmetry and the Gaussian pseudo-unitary ensemble of random matrices}
\author{ Zafar Ahmed and Sudhir R. Jain\\
{\em Nuclear Physics Division, Van de Graaff Building,}\\ 
{\em Bhabha Atomic Research Centre, Trombay, Mumbai 400 085, India}} 

\date{}
\maketitle

\begin{abstract}
Employing the currently discussed  notion of pseudo-hermiticity, 
we define a  pseudo-unitary group. Further, we 
develop a  random matrix theory which is invariant under such a 
group and call this  ensemble of pseudo-Hermitian random matrices as the 
pseudo-unitary ensemble. We obtain exact results for the nearest-neighbour 
level spacing distribution for (2 $\times$ 2)-matrices which has a novel form,   $s \log \frac{1}{s}$ near zero spacing. This shows  a level repulsion in marked distinction with an algebraic 
form $s^{\beta}$ in the Wigner surmise. We believe that this paves way for 
a description of varied phenomena in two-dimensional statistical mechanics, 
quantum chromodynamics, and so on.   
\end{abstract}
\vskip 0.5 truecm
\noindent
PACS Nos : 05.45.+b, 03.65.Ge
\vskip 0.5 truecm

Postulates of quantum theory require the observables to be represented by 
Hermitian operators as only real eigenvalues correspond to measurements. 
However, it has recently been emphasized that there are certain Hamiltonians 
describing the quantum systems which possess real eigenvalues even though 
they are not Hermitian. Many of these systems are invariant under space-time 
reflection, i.e. invaiant under a joint action of parity (${\cal P}$) and 
time-reversal (${\cal T}$) \cite{bender,znojil,ahmed}. In this context, the 
concept of pseudo-hermiticity was introduced \cite{mostafa} where it was shown 
that ${\cal PT}$-symmetry is a special case of pseudo-hermiticity. 
Pseudo-hermiticity of an operator or a matrix {\helv O} is simply 
defined through the condition : 
{\helv O}$^{\dagger}$=$\beeta${\helv O}$\beeta ^{-1}$ with $\beeta$ a metric 
and $^{\dagger}$ representing the usual adjoint or 
conjugate-transpose. Remarkably, it 
was subsequently shown that non-${\cal PT}$ invariant systems that possess real eigenvalues are also pseudo-Hermitian \cite{ahmed1}. Physical situations of 
great  interest belong to the  above  discussion. This  
includes two-dimensional statistical mechanics where parity and time-reversal are broken (preserving ${\cal PT}$) \cite{lerda,alonso,jain}, quantum chromodynamics where chiral 
ensembles are used to describe the statistical properties of lattice Dirac 
operator \cite{jjmv}, spin-rotation coupling leading to an anomalous g-value 
for muon \cite{papini}, and related fields. In this Letter, 
we present a random matrix theory which describes spectral fluctuations in systems which are 
pseudo-Hermitian and pseudo-unitarily invariant. The two aspects which are 
particularly notable are the simplicity of this novel description and the fact that this theory is natural when parity or (and) 
time-reversal is (are) violated.    

The problem of two-dimensional 
statistical mechanics is obviously connected with anyon physics and hence to 
the behavior of an electron in an Aharonov-Bohm 
medium \cite{nambu1}, i.e. a medium filled with non-quantized magnetic fluxes, reminiscent of the theory of fractional 
quantum Hall effect \cite{laughlin}. Important to note here is also another 
motivation which stems from a speculation due to Nambu that this might serve as a model for theoretical ideas like the quark confinement in a medium of 
monopoles \cite{nambu}. In this context, it is known that the spectral fluctuations of an Aharonov-Bohm billiard exhibits an interpolating behavior with respect to the strength of the flux line \cite{date}. These billiards are experimentally realized in terms of quantum dots in the presence of flux lines. It is of great interest to find an appropriate random matrix description for such 
${\cal PT}$-invariant systems. Pseudo-hermiticity appears in several 
contexts. It is instructive to note that  in the  mean-field,  RPA description 
of  nuclei \cite{ring}, the stability matrix leading to an eigenvalue problem can be checked to be  pseudo-unitary. In the context of regularization of quantum field theories, pseudo-hermiticity and the associated 
improper metric was used by Dirac \cite{dirac}, Pauli \cite{pauli}, and 
particularly by Gupta and Bleuler \cite{gupta}, and others \cite{lee}. Let us 
first establish the pseudo-unitary symmetry. 

Consider vectors {\bf x} and {\bf y} residing in a vector space ${\cal V}$ and a fixed metric $\beeta$. In this vector space, we 
define a pseudo-inner product ($\beeta$-norm), which can be written in the usual quantum 
mechanical notation as $\langle \mbox{\bf x}|\beeta \mbox{\bf y} \rangle$. 
We shall consider symmetry transformations which preserve the 
$\beeta$-norm between the vectors. We consider the Cayley form, 
{\helv D} = $e^{i\mbox{\helv G}}$ as a symmetry transformation acting on 
{\bf x, y} where {\helv G} is pseudo-Hermitian in accordance with 
$\beeta${\helv G}$\beeta ^{-1}$ = {\helv G}$^{\dagger}$.
Noting an interesting feature of {\helv D} :  
\begin{eqnarray}
\mbox{\helv D}^{\dagger} &=& e^{-i\mbox{\helv G}^{\dagger}} = 
 e^{-i\beeta \mbox{\helv G}\beeta ^{-1}}\nonumber \\ 
&=& \beeta e^{-i\mbox{\helv G}}\beeta ^{-1} 
= \beeta \mbox{\helv D}^{-1}\beeta ^{-1},
\end{eqnarray}
let us  call {\helv D} as pseudo-unitary with respect to $\beeta$. $\beeta$ 
equal to unity makes {\helv D} unitary trivially.  
To establish that {\helv D} is indeed a symmetry transformation, we need 
to show that the transformation preserves the $\beeta$-norm and a 
consistently-defined matrix element. 

Let us assume that {\bf x (y)} $\to $ {\bf x$^{\prime}$ (y$^{\prime}$)} = {\helv D}{\bf x} 
({\helv D}{\bf y}). Then, the pseudo-unitary symmetry is defined 
by preserving the pseudo-norm : 
\begin{equation}
\langle \mbox{\bf x$^{\prime}$}|\beeta \mbox{\bf y$^{\prime}$}\rangle = 
\langle \mbox{\helv D}\mbox{\bf x}|\beeta \mbox{\helv D}\mbox{\bf y}\rangle  
= \langle \mbox{\bf x}|\beeta \mbox{\bf y}\rangle .
\end{equation}     
In proving (2), use  $e^{-iG^{\dagger}}\beeta e^{iG}$ = 
$e^{-iG^{\dagger}}\beeta e^{iG}\beeta ^{-1}\beeta = 
e^{-iG^{\dagger}}e^{iG^{\dagger}}\beeta $ = $\beeta $. 
Under the same pseudo-unitary transformation, the matrix element of an 
arbitrary operator, {\helv A}, transforms as  
\begin{equation}
\langle \mbox{\bf x$^{\prime}$}|\beeta \mbox{\helv A$^{\prime}$}|
\mbox{\bf y$^{\prime}$}\rangle 
= \langle \mbox{\bf x}|\beeta \mbox{\helv A}|\mbox{\bf y}\rangle 
\end{equation}
if {\helv D A D}$^{-1}$ = {\helv A$^{\prime}$}. 

Let us now prove that pseudo-unitary matrices form a group under 
matrix multiplication. For closure, let {\helv D}$_1$ and {\helv D}$_2$ 
be two pseudo-unitary matrices. {\helv D}$_1${\helv D}$_2$ is pseudo-unitary 
because $\beeta ^{-1}$({\helv D}$_1${\helv D}$_2$)$^{\dagger}$$\beeta $ = 
$\beeta ^{-1}${\helv D}$_2^{\dagger}$ $\beeta \beeta ^{-1}${\helv D}$_1^{\dagger} \beeta $ = ({\helv D}$_1${\helv D}$_2$)$^{-1}$. It easily follows that 
{\helv D}$^{-1}$ is pseudo-unitary with respect to $\beeta$ if 
{\helv D} is pseudo-unitary  : 
$\beeta ^{-1}(e^{-i\mbox{\helv G}})^{\dagger}\beeta$ = 
$e^{i\beeta ^{-1}\mbox{\helv G}^{\dagger}\beeta}$ = $e^{i\mbox{\helv G}}$. 
The identity matrix acts as the unit element of the symmetry transformation. 
Finally, since the associativity is guaranteed, the 
N $\times$ N pseudo-unitary matrices form a pseudo-unitary group of order 
N, $PU(N)$. 

In the following, to keep the proceedings simple and explicit, we consider 
Hamiltonians in their matrix representations. Also, in the spirit of the 
original work of Wigner \cite{mehta}, we consider (2 $\times$ 2) matrices as they bring out most of the essence. In this context, there i a recent 
generalization of Wigner surmise for 2 $\times $ 2 matrices \cite{nadya}. 
Thus, 
we concentrate on $PU(2)$ and  consider the following pseudo-Hermitian  
matrix,  
\begin{eqnarray}
\mbox{\helv H} = \{\mbox{\helv H}_{ij}\}&= \left[\begin{array}{cc}a&-ib\\ic&a\end{array}\right],
\end{eqnarray}      
$a, b, c$ being real. Consequently, $e^{i\mbox{\helv H}}$ will be a 
pseudo-unitary matrix. For the above matrices, a metric is  
\begin{eqnarray}
\bdelta&=\left[\begin{array}{cc}0&-1\\1&0\end{array}\right]. 
\end{eqnarray}  
This metric may be interpreted as the parity operator ${\cal P}$, and the 
complex conjugation, ${\cal K}_0$ as time-reversal operator ${\cal T}$. With these operations, it 
may be verified that {\helv H} is ${\cal PT}$-invariant in addition to 
being ${\cal P}$-pseudo-Hermitian. Besides these commuting ${\cal P}$ and 
${\cal T}$ operators, if we choose ${\cal T}$ as the Pauli matrix times the 
complex conjugation, 
$\sigma _x {\cal K}_0$, they do not commute, however preserving other conclusions. 

This group admits three generators and an identity, viz., 
\begin{eqnarray}
\brho _1&=\left[\begin{array}{cc}1&0\\i&-1\end{array}\right], 
\brho _2&=\left[\begin{array}{cc}1&-i\\0&-1\end{array}\right],\nonumber \\
\brho _3&=\left[\begin{array}{cc}-1&0\\0&1\end{array}\right], 
{\bf I}&=\left[\begin{array}{cc}1&0\\0&1\end{array}\right].
\end{eqnarray}  
Note that {\helv H} = $a{\bf I} + b\brho _1$ + $ c\brho _2$ - $(b+c)\brho _3$.
It is interesting to see that 
$\brho _1$ and $\brho _2$ are pseudo-hermitian and pseudo-unitary, possessing 
eigenvalues $\pm 1$. It may be recalled that the Pauli matrices 
$\sigma _x$ and $\sigma _y$ are Hermitian and unitary. 
Further, the generators satisfy the following important properties,
\begin{eqnarray}
&&\brho _1^2 = \brho _2^2 = \brho _3^2 = {\bf I}, \nonumber \\
&&\left[\brho _i, \brho _j\right] = \sum_k C_{ij}^k\brho _k, 
\end{eqnarray} 
with $C_{12}^1=C_{12}^2=C_{23}^2=C_{23}^3=C_{31}^1=C_{31}^3=2$ and 
$C_{12}^3=5$. All the structure constants can be found with the help of 
commutation relations and symmetry properties, and they turn out to be 
$\pm 5$, $\pm 2$, or 0. Interestingly, the following 
relations between the structure constants hold : 
\begin{eqnarray}
&&C_{kl}^j = - C_{lk}^j \nonumber \\
&&\sum_{m=1}^3 \left[C_{kl}^m C_{jm}^s + C_{lj}^m C_{km}^s + 
C_{jk}^m C_{lm}^s\right] = 0,
\end{eqnarray}
thus making it a Lie group and defining a Lie algebra \cite{joshi}.

We now consider a Hamiltonian {\helv H} which is diagonalizable by 
{\helv D}, i.e., 
\begin{equation} 
\mbox{\helv H} = \mbox{\helv D}.
\left[\begin{array}{cc}E_+&0\\0&E_-\end{array}\right].\mbox{\helv D}^{-1}.
\end{equation}
The eigenvalues of {\helv H} are $a \pm \sqrt{bc}$ 
($bc \geq 0$). The corresponding matrix, {\helv D}, 
\begin{eqnarray}
\mbox{\helv D}&=\left[\begin{array}{cc}1&i/r\\ir&1
\end{array}\right],
\end{eqnarray}   
is pseudo-unitary under the metric, 
\begin{eqnarray}
\beeta&=\left[\begin{array}{cc}0&1\\1&0\end{array}\right]. 
\end{eqnarray} 

The eigenvalues are 
\begin{equation}
E_{\pm} = a \pm \left[\frac{c}{2r} + \frac{br}{2} \right]
\end{equation}
where $r = \sqrt{c/b}$ ($0 \leq r \leq \infty $). 

Consider that the matrix {\helv H} is drawn from an ensemble 
of random matrices with a Gaussian distribution given by \cite{mehta}
\begin{equation}
P(\mbox{\helv H}) = {\cal N} e^{- \frac{1}{2\sigma ^2}~tr~\mbox{\helv H}^{\dagger}
\mbox{\helv H}}. 
\end{equation}
Accordingly, the joint probability distribution of $a, b, c$ is
\begin{equation}    
P(a, b, c) = {1 \over 2 (\pi \sigma ^2)^{\frac{3}{2}}} e^{-{1 \over 2\sigma ^2} 
\left[2a^2 + b^2 + c^2 \right]}. 
\end{equation}               

From (4) and (9), we have the following relations :
\begin{equation}    
a = \frac{E_+ + E_-}{2}, ~b = \frac{E_+ - E_-}{2r}, 
~c = \frac{r(E_+ - E_-)}{2}.
\end{equation}               
The Jacobian, $J$ connecting $(a, b, c)$ and $(E_+, E_-, r)$ is 
$\frac{|E_+ - E_-|}{2r}$. With these, the joint probability distribution 
function (j.p.d.f.) of eigenvalues is 
\begin{equation}    
P(E_+, E_-) = \frac{|E_+ - E_-|}{2 (\pi \sigma ^2)^{\frac{3}{2}}} 
K_0\left(\frac{(E_+ - E_-)^2}{4\sigma ^2}\right) 
e^{-\frac{(E_+ + E_-)^2}{4\sigma ^2}}. 
\end{equation}     
Following the Dyson Coulomb gas analogy, this j.p.d.f. can be written as 
an equilibrium distribution of two interacting particles  with a partition 
function $P(E_+, E_-) \to {\cal Z}(x_1, x_2) = e^{-\beta {\cal H}(x_1, x_2)}$. 
It is interesting to note that the ${\cal H}$ has a potential term involving 
the logarithm of the modified Bessel function along with the familiar 
harmonic confinement and the two-dimensional Coulomb potential. $4\sigma ^2$ 
plays the role of inverse scaled temperature. 

Integrating with respect to $E_-$ gives the average density, shown in Fig. 1. 
This is not amenable to an analytically closed form.

Perhaps the most well-studied characterizer is the nearest-neighbour level 
spacing distribution, $P(S)$. This gives the frequency with which a certain 
spacing between adjacent levels occurs. For the Wigner-Dyson ensembles, 
$P(S) \sim S^{\beta _0}e^{-\gamma S^2}$ where $\beta _0$ is 1, 2, and 4 for 
the orthogonal, unitary, and symplectic ensembles. A wide variety of 
systems display universal properties possessed by random matrix ensembles 
as can be seen in \cite{mehta,haake,zelevinsky}. However, there are systems 
that display intermediate statistics \cite{parab,gremaud,bgs}. These 
systems range from examples of billiards in polygonal enclosures, 
three-dimensional Anderson model at the metal-insulator transition point, 
and so on. On the other hand, there have been important developments 
on non-Hermitian ensembles since long where the eigenvalues are complex 
\cite{mehta,zelevinsky}, and where an ensemble of unstable states are 
considered \cite{ullah}. 
Clearly, the ensemble developed here does not 
fall into any of the known categories and, indeed, displays some novel 
features as shown below. 

The spacing distribution, $P(S)$, is given in terms of the j.p.d.f. by 
\begin{eqnarray}
P(S) &=& \int_{-\infty}^{\infty}\int_{-\infty}^{\infty}P(E_+, E_-)
\delta (S - |E_+ - E_-|)dE_+dE_- \nonumber \\
&=& \frac{|S|}{\pi\sigma ^2}K_0\left(\frac{S^2}{4\sigma ^2}\right).
\end{eqnarray}  
This result is distincty different and very interesting (Fig. 2), 
particularly 
for its behaviour near zero spacing. Near $S = 0$, the probability 
distribution varies as $S\log {1 \over S}$. This follows from the 
asymptotic properties of the modified Bessel function.   

We present the 
form of two-time correlation function for a complex system with 
spectral properties given by Gaussian pseudo-unitary ensembles (GPUE). 
In this context, we consider a system with a Hamiltonian {\helv H} 
$\in $ GPUE, and an observable given by an operator, {\helv V} $\in$ 
an another GPUE. Imagine the system to be in thermodynamic equilibrium 
with a canonical density matrix, 
$\brho = e^{-\beta \mbox{\helv H}}$. Following 
an extensive study along the lines of \cite{srjpg}, the time correlation 
function, $C(t) = Z^{-1}(\beta)\mbox{~tr~}\brho \mbox{\helv V}(0)
\mbox{\helv V}(t)$ decays over long times $\sim (\frac{\log 1/t}{t})^3$.  

Finally, we wish to point out an aspect of general importance, 
encountered on many occasions in many-body theory. To give one concrete 
example, in the theory of collective excitations of Fermionic systems, 
a mean-field description is used where a collective state is first 
expressed in terms of particle-hole excitations \cite{ring}. 
Here, one generally 
encounters a matrix equation like {\helv H}$\Psi = \lambda \Phi$ 
with {\helv H} a Hermitian or unitary  operator. 
The above 
problem may be transformed into an eigenvalue problem for 
{\helv H}$^{\prime}$, i.e.,  {\helv H}$^{\prime }\Phi = \lambda \Phi$ 
with {\helv H}$^{\prime}$ a pseudo-Hermitian or pseudo-unitary operator. 
With this, there are many results immediately possible. First of all, 
the eigenvalues will either be real, complex-conjugate-pairs, 
unimodular, or they occur in pairs such that 
product of eigenvalues  is unimodular \cite{ahmed2}. 
Secondly, the statistical properties of the eigenvalues related 
to collective excitations will be 
distributed in accordance with the results obtained for GPUE above. 
Thirdly, there will be long-time tail for relaxation in the same spirit 
as the time correlation. 

The above results are found for 2 $\times$ 2 matrices. Although 
for N $\times$ N 
matrices, invariant under $PU(N)$, the results are not known, we 
conjecture that the fluctuation properties will have a similar 
form as above. As discussed earlier, the result found here gives 
a new universality corresponding to systems which are pseudo-unitarily 
invariant. In such systems, parity and time-reversal may be individually 
broken, preserving their joint action. This universality also includes 
those pseudo-Hermitian quantum systems where ${\cal PT}$ is broken. The 
examples discussed include quantum chromodynamics, two-dimensional 
statistical mechanics, and so on.

\newpage 

\noindent
{\bf FIGURE CAPTIONS}

\vskip 0.5 truecm

\begin{enumerate}
\item The average level density of an ensemble of 2 $\times$ 2 
random Gaussian pseudo-unitary ensemble is shown here. 

\item The nearest-neighbor level-spacing distribution is shown 
here. For comparison, the results corresponding to the Wigner-Dyson 
ensembles corresponding to orthogonal and unitary symmetries  are also 
shown. Whereas the level repulsion is linear and quadratic in the 
orthogonal and unitary ensembles, here it is of the form $s \log (1/s)$,   
as shown in the inset. This then is a new universality.  
\end{enumerate}
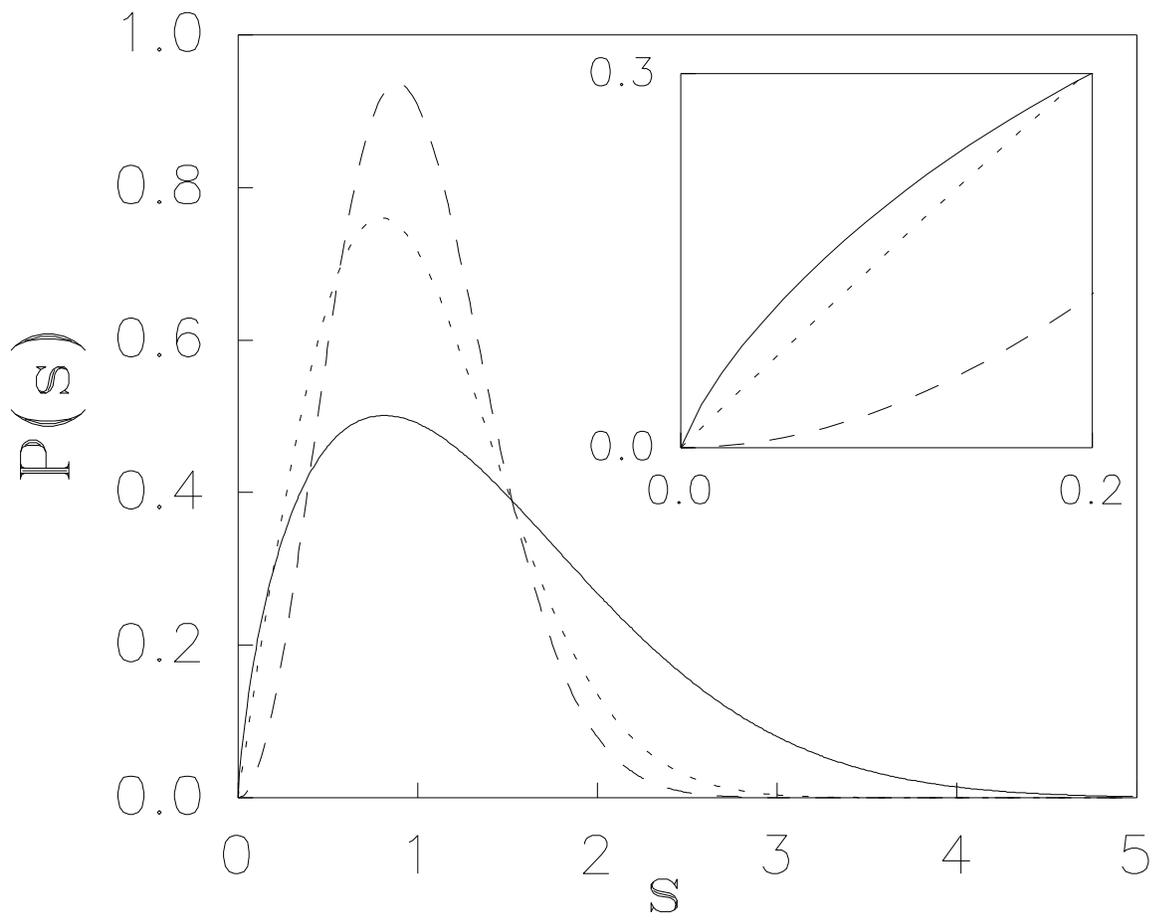
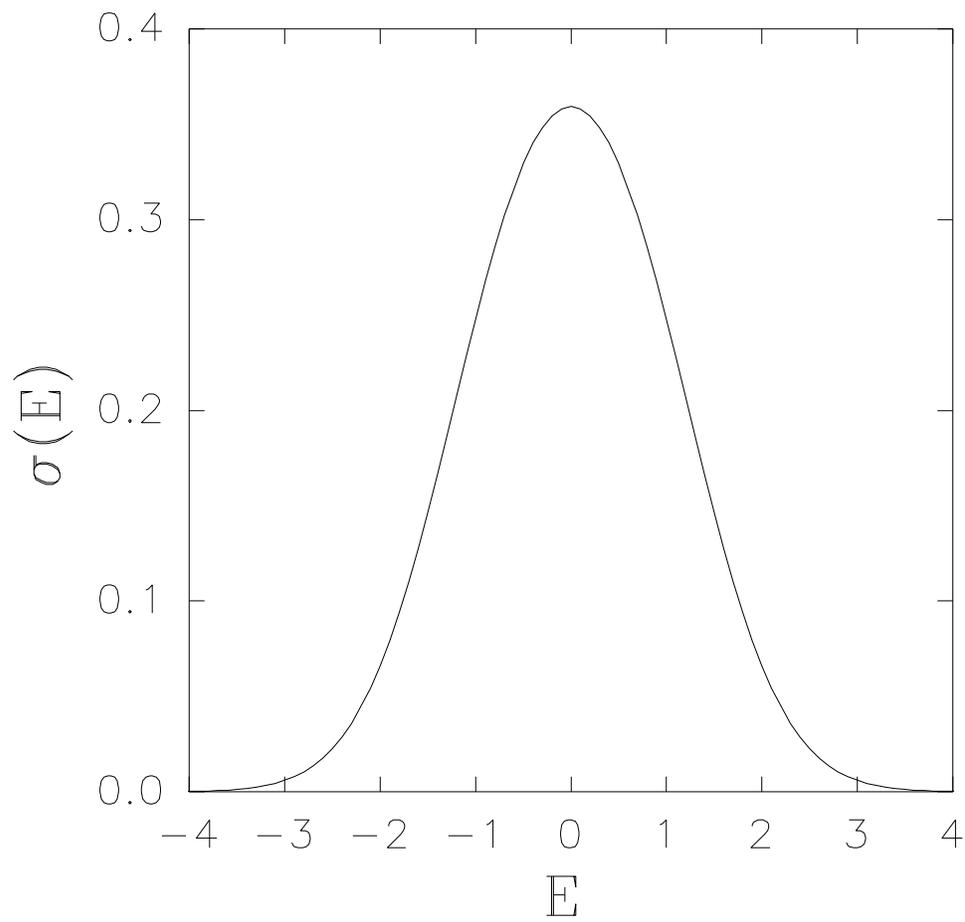
\end{document}